\documentclass[aps,reprint,twocolumn,superscriptaddress,10pt]{revtex4-1}
\usepackage{siunitx}
\usepackage{amsfonts}
\usepackage{amsmath}
\usepackage{amssymb}
\usepackage{graphicx}
\usepackage{dcolumn}
\usepackage{mciteplus}
\usepackage{euscript}
\usepackage{multirow}
\usepackage{color}
\usepackage{textcomp}
\usepackage{gensymb}
\providecommand{\U}[1]{\protect\rule{.1in}{.1in}}

\def\ms{M_\text{s}}
\def\nt{N_\perp}
\def\nll{N_\parallel}
\newcommand{\wfig}{0.44\textwidth}
\newcommand{\ala}{\alpha_1}
\newcommand{\alb}{\alpha_2}

\newcommand{\hc}{H_\text{c}}

\newcommand{\redbf}[1]{}

\newcommand{\memome}[1]{}
\newcommand{\req}[1]{Eq.~(\ref{#1})}

\newcommand{\rfig}[1]{Fig.~\ref{#1}}
\newcommand{\rFig}[1]{Figure~\ref{#1}}

\newcommand{\st}{\text{ST}}
\newcommand{\bct}{\text{BCT}}

\begin{document}
\title{Simulation of alnico coercivity}

\author{Liqin Ke}
\email[Corresponding author: ]{liqinke@ameslab.gov}
\affiliation{Ames Laboratory, U.S. Department of Energy, Ames, Iowa 50011, USA}
\author{Ralph Skomski}
\affiliation{Department of Physics and Astronomy, University of Nebraska, Lincoln, Nebraska 68588, USA}
\author{Todd D. Hoffmann}
\affiliation{1137 W Emerald Ave, Mesa, Arizona 85210, USA}
\author{Lin Zhou}
\affiliation{Ames Laboratory, U.S. Department of Energy, Ames, Iowa 50011, USA}
\author{Wei Tang}
\affiliation{Ames Laboratory, U.S. Department of Energy, Ames, Iowa 50011, USA}
\author{Duane D. Johnson}
\affiliation{Ames Laboratory, U.S. Department of Energy, Ames, Iowa 50011, USA}
\affiliation{Department of Materials Science \& Engineering, Iowa State University, Ames, Iowa 50011, USA}
\author{Matthew J. Kramer}
\affiliation{Ames Laboratory, U.S. Department of Energy, Ames, Iowa 50011, USA}
\author{Iver E. Anderson}
\affiliation{Ames Laboratory, U.S. Department of Energy, Ames, Iowa 50011, USA}
\author{C.-Z. Wang}
\affiliation{Ames Laboratory, U.S. Department of Energy, Ames, Iowa 50011, USA}

\begin{abstract}
Micromagnetic simulations of alnico show substantial deviations from
Stoner-Wohlfarth behavior due to the unique size and spatial
distribution of the rod-like Fe-Co phase formed during spinodal
decomposition in an external magnetic field. The maximum coercivity is
limited by single-rod effects, especially deviations from ellipsoidal
shape, and by interactions between the rods. Both the exchange
interaction between connected rods and magnetostatic interaction
between rods are considered, and the results of our calculations show
good agreement with recent experiments. Unlike systems dominated by
magnetocrystalline anisotropy, coercivity in alnico is highly
dependent on size, shape, and geometric distribution of the Fe-Co
phase, all factors that can be tuned with appropriate chemistry and
thermal-magnetic annealing.
\end{abstract}

\date{\today}
\maketitle

The anisotropy of most permanent magnets is of magnetocrystalline
origin, meaning that hysteresis and coercivity rely on the
atomic-scale interplay between spin-orbit coupling and crystal-field
interaction~\cite{herbst.rmp1991,kumar.jap1988,skomski.book1999}.
Alnico magnets---a family of nanostructured alloys consisting
primarily of Fe, Al, Ni, and Co---are an exception, because their
magnetic anisotropy and hysteresis originate almost entirely from
magnetostatic dipole-dipole interactions
~\cite{mishima.ohm1932,bozorth.book1951,mccurrie1982coll-c3sp,zhou.acta2014}.
These materials have attracted renewed attention in the context of
magnetic materials that are free of rare-earth elements and do not
contain other expensive elements, such as
Pt~\cite{zhou.acta2014,jones.nat2011,gutfleisch.am2011,coey.scma2012,chu.nat2012,mccallum.arms2014,skomski.ieetm2013,zhou.mmte2014}. The
magnetic anisotropy of alnicos reflects their peculiar nanostructure,
where high-magnetization rods with an approximate composition of FeCo
($\ala$-phase) are embedded in an essentially nonmagnetic Al-Ni-rich
matrix
($\alb$-phase)~\cite{mishima.ohm1932,bozorth.book1951,mccurrie1982coll-c3sp,zhou.acta2014,zhou.mmte2014,sun.jmmm2015,zou.jmmm2016,lowe.jmmm2016,fan.ieeetm2016}.

\begin{figure}[hbt]
\centering
\begin{tabular}{c}
\includegraphics[width=0.58\linewidth,clip]{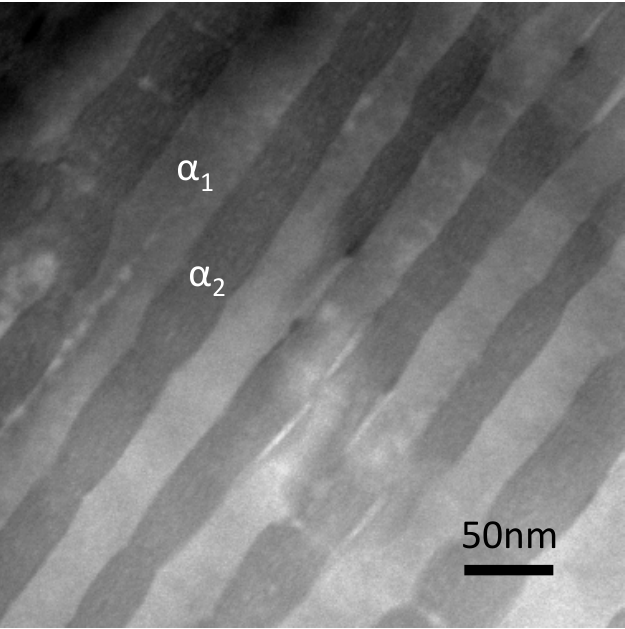}
\end{tabular}%
\caption{HAADF STEM image showing a side view of $\ala$ rods
  distributed in the $\alb$ matrix in an alnico 8 alloy.}
\label{fig:alnico_tem}
\end{figure}

There are several grades of alnico magnets, characterized by different
chemical compositions and microstructures. We focus on a high quality
grade, namely, alnico 8. \rFig{fig:alnico_tem} shows a
high-angle-annular-dark-field (HAADF) scanning transmission electron
microscopy (STEM) image of an alnico 8 sample along the longitudinal
direction. The $\ala$ rods in the sample of \rfig{fig:alnico_tem} have
diameters of $\sim$\SIrange{25}{45}{\nm} and lengths of
$\sim$\SIrange{100}{600}{\nm} and are uniformly distributed in the
$\alb$ matrix. Some of the $\ala$ rods have pointy ends and/or touch
each other. The detailed microstructures strongly depend on alloy
composition and heat treatment conditions. Details of alnico alloy
fabrication and microstructure characterization are reported
elsewhere~\cite{zhou.acta2014,zhou2017am}.

Alnico magnets have high Curie temperatures and magnetizations, but
their modest coercivity limits the performance of this otherwise very
good permanent-magnet material. Surprisingly, the understanding of
alnico coercivity in terms of the dipolar anisotropy has remained very
poor quantitatively and even qualitatively. The main reason is the
multiscale character of the calculations, which involves local
features having sizes of less than \SI{5}{\nm}, but ranging to
interactions on scales comparable to or exceeding the wire length of
about \SI{1}{\um}. Only recently, computer power has become sufficient
to treat interactions between the rods. In this letter, we present
micromagnetic simulations~\cite{fidler.jpdap2000} to quantitatively
explain the coercivity of alnico magnets. In particular, we show and
analyze how the coercivity depends on structural features, namely, the
shape of the rod ends, the spatial arrangements of Fe-Co rods in the
magnet, and the crossing and/or branching of rods.

The conventional explanation of alnico anisotropy is shape anisotropy
similar to that of small, elongated Stoner-Wohlfarth
particles~\cite{luborsky.jap1957,chikazumi.book1964}. In this
approximation, an aligned magnetic rod or ``elongated fine particle''
is subject to a mean-field-like interaction with neighboring rods, and
the corresponding coercivity is given by the semi-empirical formula
~\cite{luborsky.jap1957}
\begin{equation}
  \hc=(1-p)(\nt-\nll)\ms.
\label{eq:01}
\end{equation}
Here, $\nt\approx 1/2$ and $\nll\approx 0$ are the demagnetization
factors perpendicular and parallel to each rod's long axis, $\ms$ is
the magnetization of the rods, and $p$ is their packing fraction in
the nonmagnetic matrix. However, the coherent-rotation or
Stoner-Wohlfarth model has several limitations.

First, it is limited to rods of very small diameters, less than
$2R_\text{coh}\approx\SI{20}{\nm}$
~\cite{brown.rmp1945,zeng.prb2002,skomski.jpcm2003}, whereas typical
alnico rods have diameters of the order of $2R=\SIrange{40}{50}{\nm}$
(\rfig{fig:alnico_tem}). In such relatively thick rods, the
magnetization reversal starts by magnetization curling, for which the
nucleation field is described by
~\cite{mccurrie1982coll-c3sp,brown.rmp1945,zeng.prb2002,skomski.jpcm2003,skomski.book2008,aharoni.book1996}
\begin{equation}
  \hc=\frac{2K_1}{\mu_0\ms}-\nll\ms+\frac{c(\nll)A}{\mu_0\ms R^2}.
\label{eq:02}
\end{equation}
Here, $K_1$ is the magnetocrystalline anisotropy constant, $A$ is the
exchange stiffness~\cite{donahue.pbcm1997}, and the values of $c$ are
8.666 for spheres ($\nll=1/3$) and 6.678 for needles ($\nll=0$). In
alnicos, the $K_1$ term is small compared to the magnetostatic terms,
contributing only about \SI{10}{\percent} to $\hc$
~\cite{iwama.tjim1976}, and usually neglected
~\cite{skomski.ieetm2013,luborsky.jap1957}.  In \req{eq:01}, the
difference $\nt-\nll=1-3\nll$ is positive for $\nll<1/3$ (shape
anisotropy), but the corresponding curling term $-\nll\ms$ in
\req{eq:02} is always negative. The last term in \req{eq:02} partially
compensates the coercivity loss due to curling but depends on the
radius $R$. Since alnico rods are thick enough that reversal starts
via curling, it is simplistic to interpret alnico anisotropy as shape
anisotropy~\cite{skomski.jpcm2003,skomski.aipadv2017}.  In reality,
$-\nll\ms$ is negative but small enough that the exchange term in
\req{eq:02} ensures a positive coercivity unless the rod diameter $2R$
is very large.  In practice, the diameter\textemdash determined by the
spinodal decomposition process\textemdash is sufficiently small to
contribute some coercivity but not small enough to reach the
Stoner-Wohlfarth limit of \req{eq:01}.

Second, Eqs.~(\ref{eq:01}--\ref{eq:02}) are only valid for
ellipsoids. For real structures other than ellipsoidal shapes, $\nt$
and $\nll$ are no longer well-defined, and the non-ellipsoidal shapes
have smaller coercivity than the ellipsoidal
shape~\cite{zeng.prb2002,skomski.jpcm2003,fischbacher.repm2014}. This
complication, commonly referred to as Brown's paradox, can be resolved
only by explicit consideration of the magnet's real structure (bulk
microstructure).

Third, interactions between rods are simplistically treated in
\req{eq:01} and formally ignored in \req{eq:02}. To be precise,
\req{eq:01} replaces the complicated magnetostatic interaction by a
$p$-dependent mean field. The coercivity is the largest for distantly
spaced rods ($p\approx0$) but vanishes completely in the limit of
continuous soft-magnetic thin films ($p=1$). The energy product, which
succinctly expresses the performance of a permanent magnet, reaches
its maximum at $p=2/3$ in this
approximation~\cite{skomski.ieetm2013,skomski.jap2010}, agreeing
fairly well with experiment. The same magnetostatic interaction effect
could be included in \req{eq:02} by introducing effective
demagnetizing factors ~\cite{skomski.ieetm2007}, but this does not
alleviate the basic shortcomings of the approximation. In fact, the
coercivity strongly depends on the spatial arrangement of the magnetic
rods, as we will see below.

Fourth, it is known experimentally that the rods or ``wires'' of the
magnetic phase undergo crossing and branching
~\cite{mccurrie1982coll-c3sp,zhou.acta2014,zhou.mmte2014}. These
features are likely to strongly affect coercivity, and their treatment
requires demanding numerical calculations. In the theoretical
literature, this effect has not been addressed so far.

To simulate the coercivity, we model the alnicos as aligned or
``anisotropic'' magnets consisting of parallel magnetic rods, thereby
establishing a unique $\parallel$ (parallel) axis.  This structural
model reproduces the key feature of alnico microstructure, namely,
magnetic rods in an essentially nonmagnetic matrix
(\rfig{fig:alnico_tem}). Some alnicos have rod orientations that
differ from the global magnetization axis, i.e., the rods are
misaligned ~\cite{mccurrie1982coll-c3sp,zhou.acta2014,zhou.mmte2014}.
This case is physically very different~\cite{skomski.jap2014} from the
presently aligned rods and goes beyond the scope of the present paper.

Our calculations use the recently developed \textsc{MuMax3}
micromagnetics code ~\cite{vansteenkiste.aipadv2014}.  We employed
either clusters of particles or periodic boundary conditions (PBCs) to
investigate the interactions between rods.  In our calculations, we
used a \SI{1}{\nm} grid and \SI{1}{Oe} field steps to produce the
hysteresis loops. At each field step, we computed equilibrium magnetic
states by directly minimizing energy using the steepest descent
method~\cite{vansteenkiste.aipadv2014,exl2014jap} as implemented in
\textsc{MuMax3}. To simulate the (Fe-Co)-rich magnetic phase, we have
assumed a saturation magnetization of $\mu_0\ms=\SI{2.1}{\tesla}$ and
exchange stiffness of
$A=\SI[per-mode=symbol]{11}{\pico\joule\per\meter}$.  To explore the
coercivity's sensitivity to these inputs, we also varied $\ms$ and $A$
and found that $\hc$ only changes a small amount if the ratio $A/\ms$
remains constant.  For example, doubling both $\ms$ and $A$ (for an
ellipsoid with $D=\SI{32}{\nm}$ and $c/a=8$) yields a coercivity
difference of less than \SI{10}{\percent}. Finite-temperature dynamic
effects have not been considered explicitly, because they are known to
yield well-understood logarithmically small magnetic-viscosity
corrections~\cite{skomski.book1999,skomski.jpcm2003}.

\begin{figure}[ht]
\centering
\begin{tabular}{c}
\includegraphics[width=0.42\textwidth,clip]{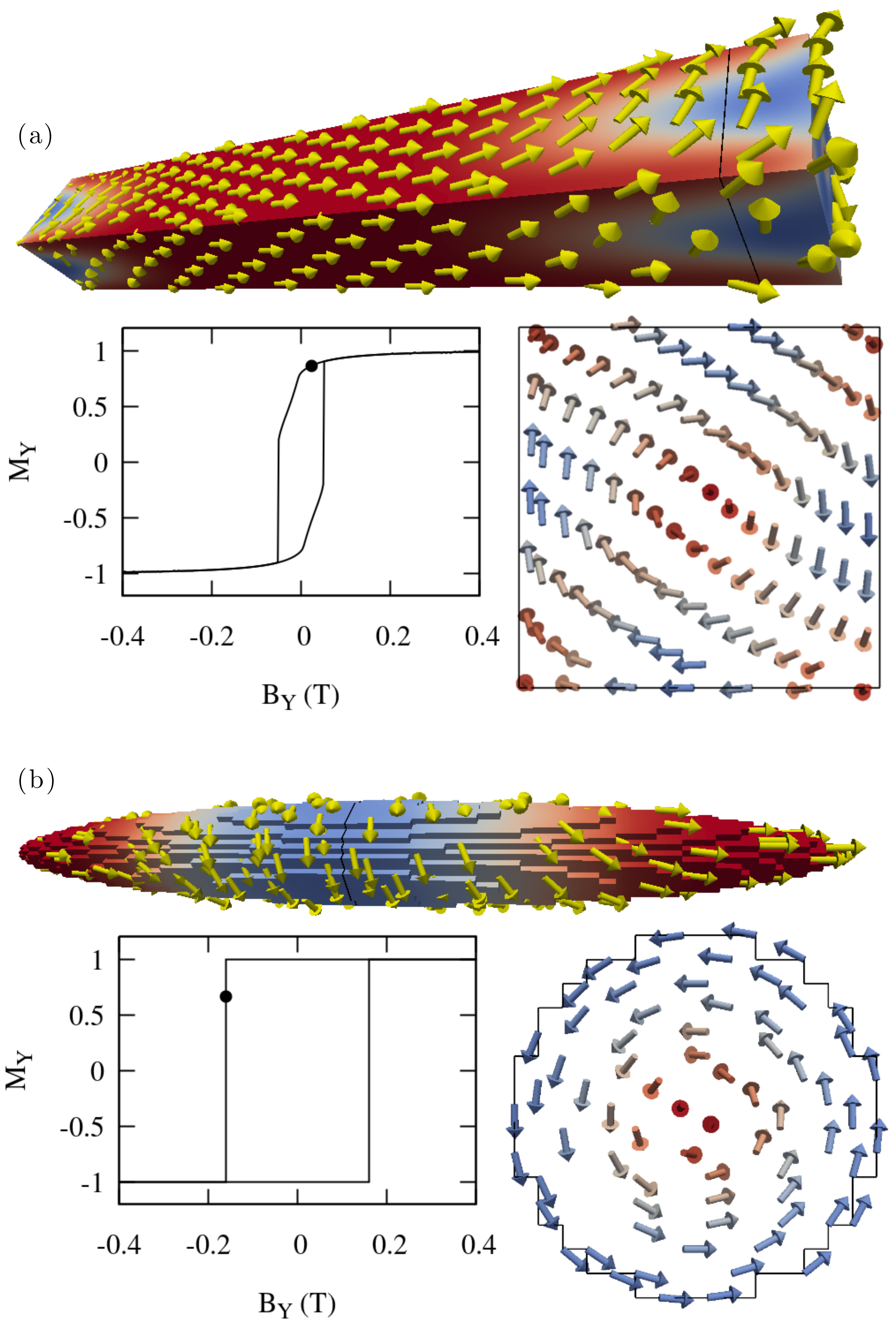}
\end{tabular}%
\caption{Spin structures at the initial stage of curling for isolated
  Fe-Co (a) square prism and (b) ellipsoid surfaces and
  cross-sections. The corresponding fields and magnetizations are
  shown on the hysteresis loops. The surface coloring visualizes the
  magnetization along the field direction.}
\label{fig:Curling_isolated_cuboid_ellipsoid}
\end{figure}

\rFig{fig:Curling_isolated_cuboid_ellipsoid} visualizes the initial
stage of magnetization reversal, which is of the curling type, and
shows the corresponding hysteresis loops. In the cuboids, the curling
starts at the ends of the rod, in the middle of the short edges,
propagates along the middle of the long faces, and eventually advances
to the long edges and center of the rod. The curling mode in the
ellipsoids is essentially delocalized throughout the rod, in agreement
with exact analytic calculations~\cite{aharoni.book1996}. Note the
nonrectangular (curved) shape of the cuboid hysteresis loops.

\begin{figure}[ht]
\centering
\begin{tabular}{c}
\includegraphics[width=\wfig,clip]{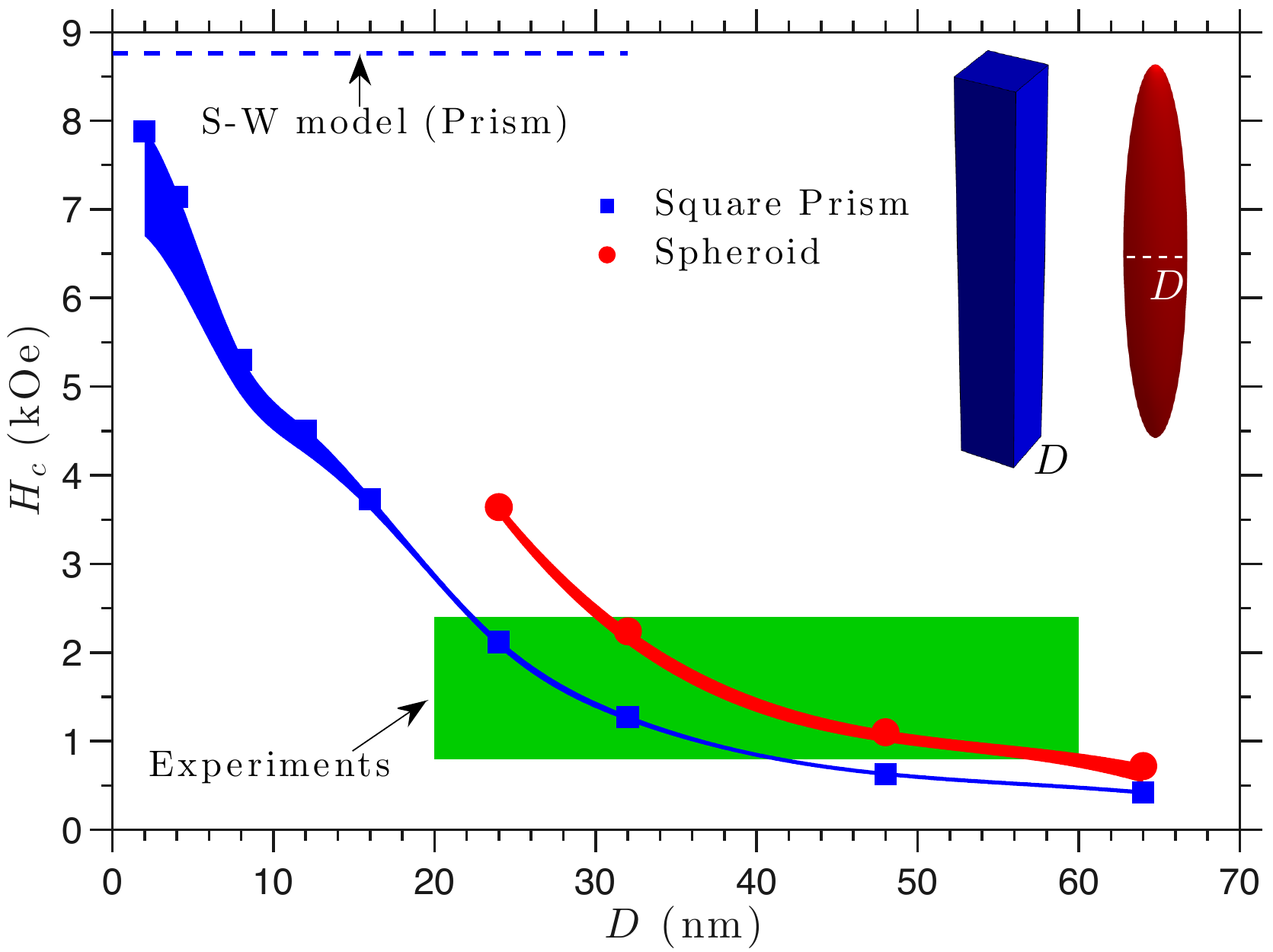}%
\end{tabular}%
\caption{Coercivity of isolated magnetic rods of aspect ratio 8 as a
  function of rod diameter. The Stoner-Wohlfarth limit for cuboidal
  rod is denoted by the dashed (blue) line. The thickness of the blue
  and red solid lines indicates the coercivity variation when the
  angle between the external field and rods varies from
  \SI{1}{\degree} to \SI{5}{\degree}. Typically observed alnico Fe-Co
  rod sizes and coercivities are denoted by the shaded (green)
  region.}
\label{fig:hc_vs_size}
\end{figure}

It is straightforward to extract the coercivities from the calculated
hysteresis loops. \rFig{fig:hc_vs_size} shows the coercivities of
isolated rods of aspect ratio 8 as a function of width $D$ for the
square prism or diameter $D=2R$ for the ellipsoid. The overall
behavior, namely, that our coercivity results approach the
Stoner-Wohlfarth value for very thin rods and nearly vanish for thick
rods, is consistent with Eqs.~(\ref{eq:01}) and
(\ref{eq:02}). Furthermore, the square prism rods have smaller
coercivity than the ellipsoids, in accordance with \req{eq:01} since
the demagnetizing field is inhomogeneous in square prisms.

\begin{figure}[ht]
\centering
\begin{tabular}{c}
  \includegraphics[width=\wfig,clip]{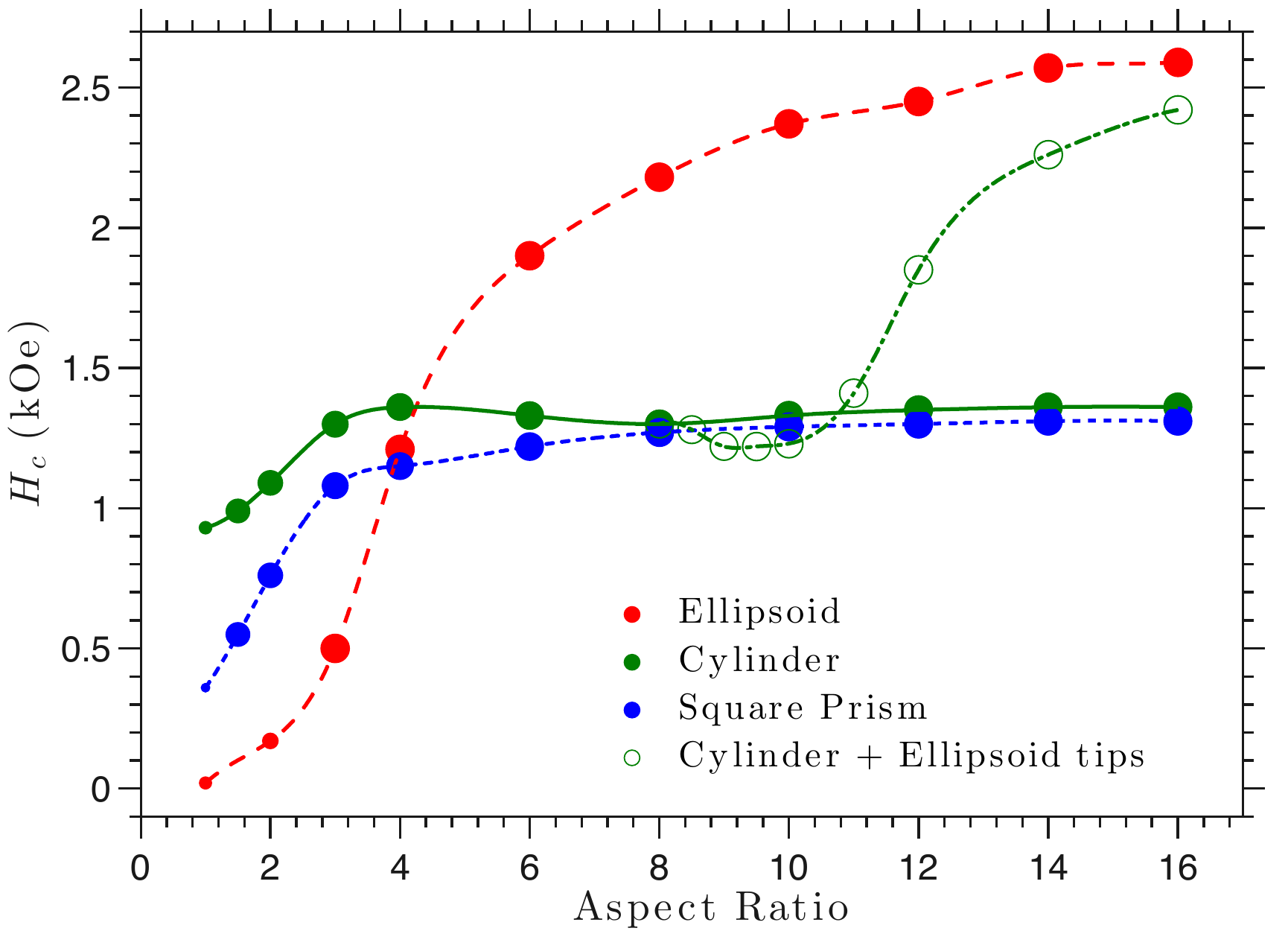}  
\end{tabular}%
\caption{Coercivity versus aspect ratio for different rod
  geometries. $D=\SI{32}{\nm}$ in all cases. The ellipsoid-capped
  cylinders are constructed by capping a 256-\si{\nm}-long cylinder
  with hemiellipsoids of various aspect ratio.}
\label{fig:hc_vs_coa_ellipsoid}
\end{figure}

\rFig{fig:hc_vs_coa_ellipsoid} analyzes the dependence of the
coercivity on the aspect ratio of the rods. For ellipsoids, $\hc$
slowly approaches a plateau value, because $\nll$ continues to change
as the rods get longer. In contrast, $\hc$ in both square prisms and
cylinders barely changes above $c/a=4$. The behavior of cylinders
capped by ellipsoidal tips is intermediate between cylinders and
ellipsoids, and the details depend on the aspect ratio of the tip,
where hemispherical tips have little impact on the coercivity, but
elongated tips improve it. This indicates that the geometry of the tip
ends is more important than the aspect ratio and that ellipsoidal tips
are (far) better than flat tips.

The green rectangle in \rfig{fig:hc_vs_size} shows the range of rod
diameters and coercivities typically encountered in laboratory-scale
and industrial practice~\cite{zhou.acta2014,mccurrie1982coll-c3sp}.
Compared to the Stoner-Wohlfarth predictions (dashed line), the
single-rod approximation of Figs.~1--3 reproduces the correct order of
magnitude for coercivity.  However, the quantitative agreement is, by
no means, perfect.  More importantly, interactions between rods are
likely to modify the coercivity in a qualitative way.  On a mean-field
level, the magnetostatic interaction corrections are approximated by
\req{eq:01}, but specific alnico interaction mechanisms have not yet
been considered in the literature.  Two classes of interactions need
to be addressed: geometry-determined magnetostatic interactions and
exchange interactions between bridging or branched rods.

\begin{figure}[th]
\centering
\begin{tabular}{c}
  \includegraphics[width=\wfig,clip]{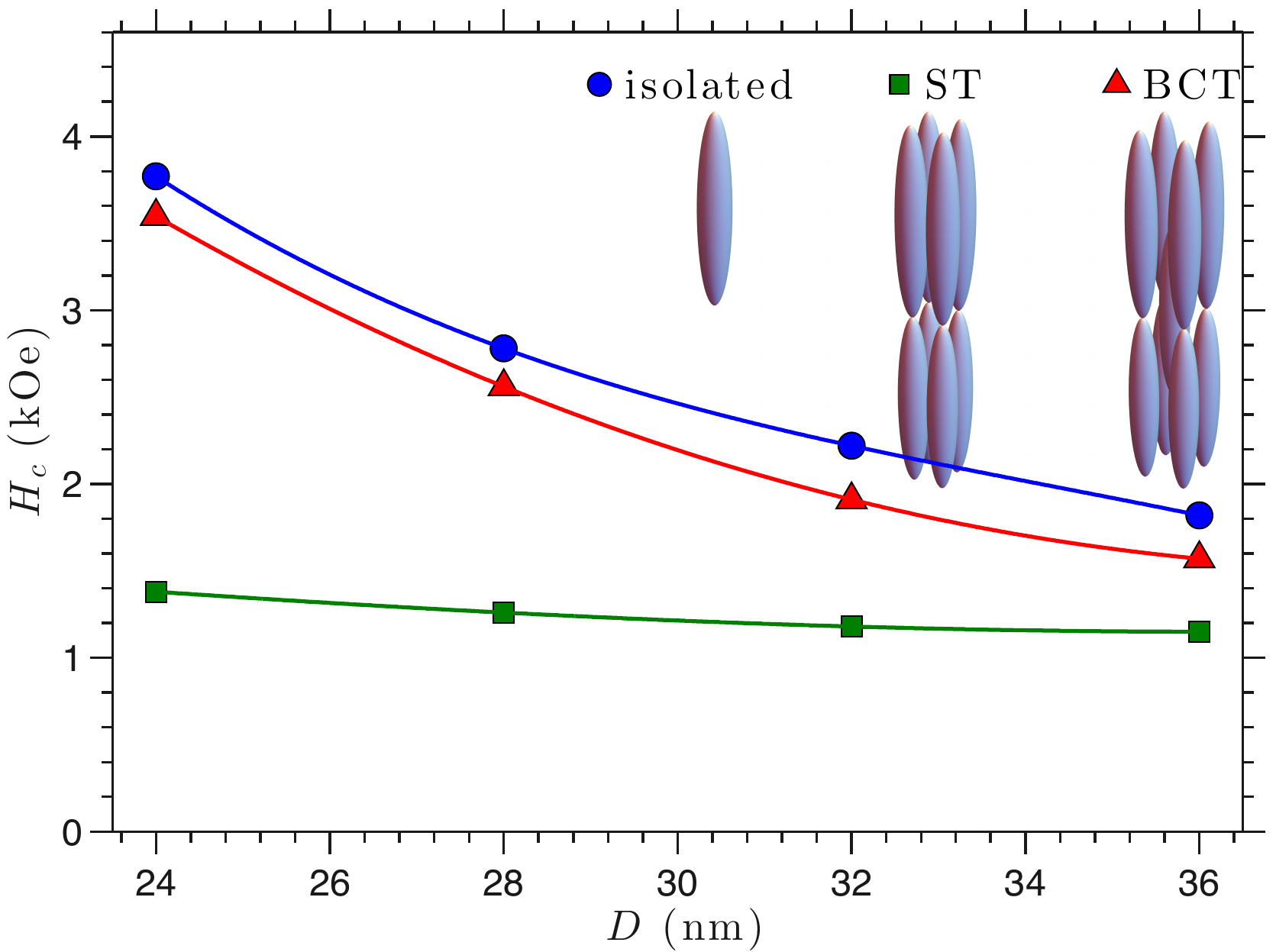} \\  
\end{tabular}%
\caption{Coercivity versus rod size for isolated ellipsoids and their
  assemblies (calculated using PBCs) with simple-tetragonal ($\st$)
  and body-centered tetragonal ($\bct$) patterns.}
\label{fig:hc_vs_pbc}
\end{figure}

To explore magnetostatic interaction effects, we have compared two
arrangements of ellipsoidal rods, namely simple tetragonal ($\st$)
rods and body-centered tetragonal ($\bct$) rods
(\rfig{fig:hc_vs_pbc}).  The $\st$ array is constructed by repeating
rods along the $x$, $y$, and $z$ directions.  The $\bct$ geometry is a
staggered array, where nearest-neighbor particles are shifted along
the $\parallel$ direction so that the center of the rods is between
the tips of its eight neighboring rods. Periodic boundary conditions
(PBCs) implemented using a so-called macrogeometry
approach~\cite{fangohr.jap2009} were applied to simulate the assembly.
\rFig{fig:hc_vs_pbc} compares the two arrangements for a constant
packing fraction of $p\approx0.5$, an aspect ratio of 8, and a
vertical end-to-end spacing of \SIrange{1}{2}{\nm}.  We see that the
coercivity of the $\bct$ array is about twice as high as the
coercivity of the $\st$ array.  The reason for the difference is that
tips with $\uparrow$ magnetization act as poles and create a strong
$\downarrow$ field in the lateral neighborhood, adding to the reverse
external magnetic field.  The $\st$ array has four laterally
coordinated nearest neighbors, whereas in the $\bct$ array, the
nearest lateral neighbors are more distant.  Furthermore, top and
bottom tips form magnetic poles of opposite sign, which further
reduces the lateral interaction effect for $\bct$ array.  For the
coercivity of square prisms calculated using PBCs, the difference
between $\st$ and $\bct$ arrays is smaller.

We have also calculated the coercivity of an {\it isolated} cluster of
64 ellipsoidal rods in a $\bct$ arrangement, and the coercivities are
very similar to those obtained by using PBCs. The $1-p$ dependence of
$\hc$ on packing fraction, estimated using mean field as in
\req{eq:01}, underestimates $\hc$ in this case.

\begin{figure}[ht]
\centering
\begin{tabular}{c}
\includegraphics[width=\wfig,clip]{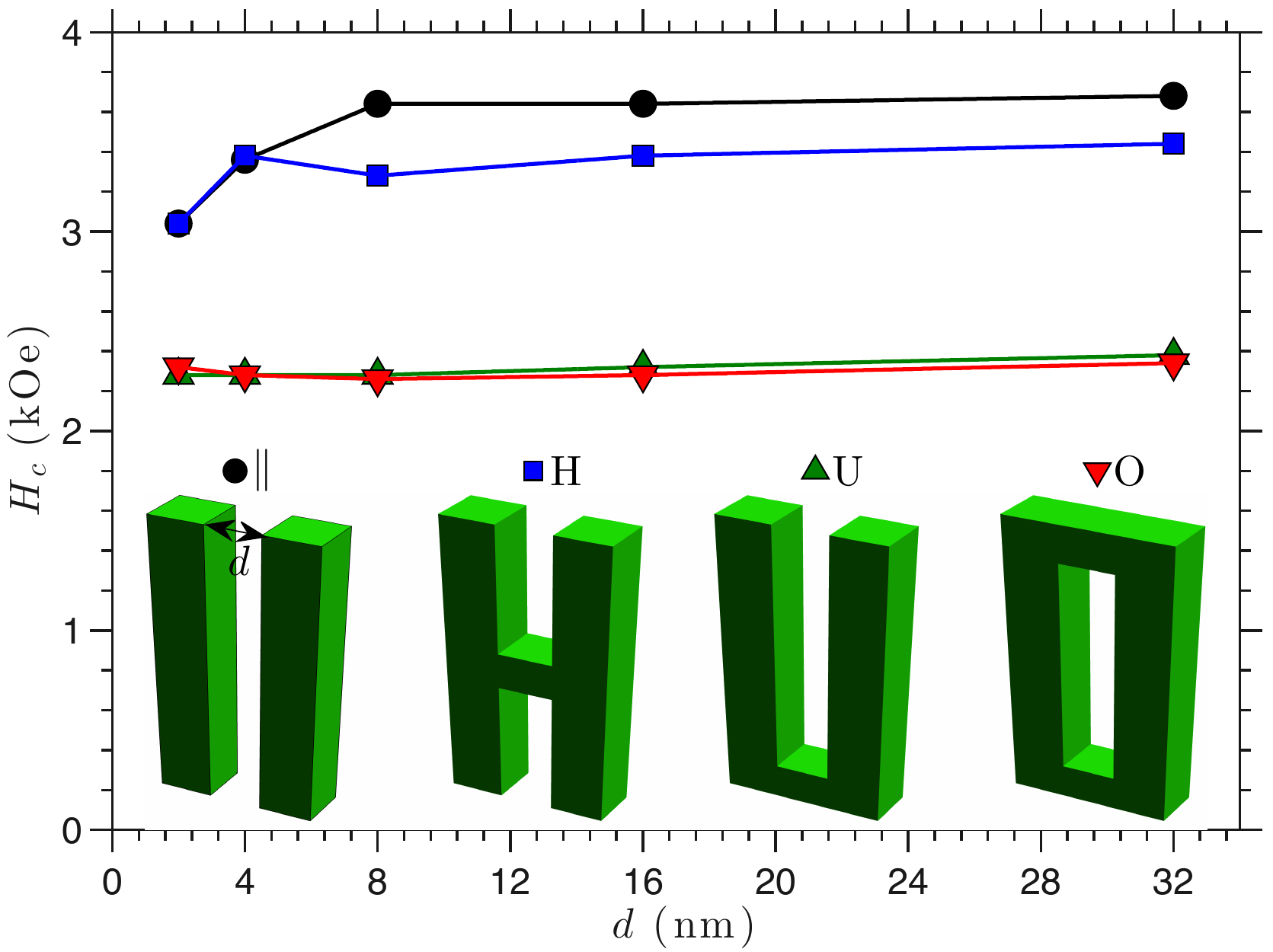}%
\end{tabular}%
\caption{Coercivity of free and connected rods as a function of the
  spacing $d$ between rods. The individual rods have a width of
  \SI{16}{\nm} and an aspect ratio value of 8.}
\label{fig:cuboid_branching}
\end{figure}

Strong exchange interactions are established through bridges and
branching between rods. We have considered two parallel cuboid
nanorods that are connected through different types of branching,
namely, H-shaped, U-shaped, and O-shaped
geometries. \rFig{fig:cuboid_branching} shows the coercivities as a
function of the surface-to-surface distance $d$ between rods. H-shape
branching, where the rods are connected in the middle, is much less
detrimental to coercivity than U- or O-shaped branching, where the
rods are connected at the ends. This is because the magnetization
reversal starts at the ends of the rods and is made easier by U- and
O-branches in the vicinity. In the H-shape geometry, the reversal also
starts at the rod ends, but the tips are still fairly isolated and the
branches in the middle have a minor effect.

Interestingly, some of the above features are also encountered in
fine-particle and nanowire magnetism
~\cite{chikazumi.book1964,zeng.prb2002,poudyal.jpdap2013,gandha.sr2014,toson.ieeetm2015,vinas.nanotech2015,niarchos.jom2015}.
For example, fine particles also exhibit a transition from coherent
rotation to curling, and wire-end features affect the coercivity
~\cite{zeng.prb2002,skomski.jpcm2003,ott.jap2009}. Fine particles and
embedded wires often have diameters small enough to approach
Stoner-Wohlfarth $\hc$ values but are nevertheless difficult to use as
permanent magnets. The reason is that with their relatively small
packing fraction $p$ the energy product of a permanent magnet, which
describes its performance~\cite{skomski.book1999}, is quadratic in the
magnetization and, therefore, quadratic in $p$. In contrast, the
nanostructure of alnico results from spinodal decomposition via
specific heat treatments and alnico packing fractions approach the
``ideal value'' of $p=2/3$, depending on alloy composition. This
yields relatively high magnetization levels but also strengthens the
interactions between rods that can reduce coercivity.

In conclusion, we have used micromagnetic simulations to analyze the
coercivity of alnico permanent magnets. We find strong deviations from
the Stoner-Wohlfarth model caused by curling modes that are modified
by the real structure of alnico alloys. Both the absolute coercivities
and the coercivity trends are in good agreement with available
experimental data. The shape, size, spacing, volume fraction,
arrangements, and branching types of the magnetic Fe-Co rods in the
nonmagnetic Ni-Al matrix all affect the coercivity, but aside from the
packing fraction, the most important features are the shape of the rod
ends or tips and interactions between them. These are all factors
which are controlled by the chemistry and thermal-magnetic annealing
of these alloys~\cite{cahn.acta1961,zhou.acta2014,zhou.mmte2014}. We
predict that sharp ellipsoidal tips and a staggered arrangement of the
rods should promote substantial coercivity improvements, but this
morphology may also be the most difficult one to realize
experimentally. Further research is necessary to see, for example,
whether field annealing can be used to realize a staggered
configuration and if optimized draw annealing can lead to rod tip
``sharpening''.

\textbf{Acknowledgment}
The authors are thankful to D. J. Sellmyer, W.-Y. Zhang, H.-W. Kwon,
and K.-M. Ho for discussing several aspects of the present paper. This
work was supported by the U.S. Department of Energy (DOE) EERE-VT-EDT
program under the DREaM project at the Ames Laboratory, which is
operated for the U.S. DOE by Iowa State University under Contract
No. DE-AC02-07CH11358.

\bibliography{aaa,methods}
\bigskip 

\end{document}